\documentclass[12pt]{article}
\usepackage{graphics}
\input epsf

\newcommand{\be}{\begin{eqnarray}}
\newcommand{\ee}{\end{eqnarray}}

\begin{document}
\title{Beyond the Zero-Binding Approximation in Quarkonium}
\author{L.  Clavelli\footnotemark[1]\\
Department of Physics and Astronomy\\ University of Alabama\\
Tuscaloosa AL 35487\\
\and
P. H.  Cox\footnotemark[2]\\
Department of Physics and Geosciences\\ Texas A\&M University-Kingsville\\
Kingsville TX 78363\\
\and
T.  Gajdosik\footnotemark[3]\\
Department of Physics and Astronomy\\ University of Alabama\\
Tuscaloosa AL 35487}
\footnotetext[1]{lclavell@bama.ua.edu}
\footnotetext[2]{phcox@tamuk.edu}
\footnotetext[3]{garfield@bama.ua.edu}
\date{October 7, 2002}
\maketitle
\begin{abstract}
The hadronic decays of quarkonium and the B meson inclusive decay into
$J/\Psi + X$, if treated in the zero-binding approximation, suggest a
value of the strong coupling constant much smaller than the value
implied in the standard model by running from measurements at the $Z$.
Thus, assuming the standard model is correct in the low energy region,
there must be very substantial relativistic binding corrections to these
processes.  We discuss the wave function factor appearing in
covariant treatments of quarkonium production and decay processes with
special attention to the way in which the spin of the quarks carries
over into the spin of the bound state.  
We find that Lorentz covariance requires that the bound state be a 
superposition of free quark and antiquark spinors notably 
different from the usual ones.
We resolve a superficial apparent paradox
suggesting that the relative momentum of the quarks should lie in a
plane perpendicular to the spin quantization axis and we calculate the
$J/\Psi$ binding corrections to lepton pair decay and to the inclusive 
B decay.

\end{abstract}
\vfill\newpage
\renewcommand{\theequation}{\thesection.\arabic{equation}}
\renewcommand{\thesection}{\arabic{section}}
\section{\bf Introduction}
\setcounter{equation}{0}
\par
 
      Although perturbative QCD was originally seen as the source of
the Zweig rule leading to the narrowness of heavy quarkonium \cite{APDG}
, measurements
of the strong coupling constant at the Z boson would imply, in the
non-relativistic or zero-binding approximation  (ZBA), hadronic decay rates
significantly greater than those observed \cite{ClavelliCoulter}.  
Thus, if the standard model
is correct in the quarkonium region, corrections to these decay rates
beyond the ZBA must greatly suppress these rates.
A similar problem exists in the treatment of the inclusive B meson decay
into $J/\Psi + X$.  
After correcting an error in \cite{Cox} noted by \cite{Bodwinetal}, the
color singlet model gives an $\alpha_s(m_b)$ some $20\%$ lower than suggested by 
extrapolating in the standard model from measurements at the $Z$ boson
(i.e., $0.154
\pm 0.005$ versus $0.20 \pm 0.01$).
These low values of the strong coupling constant
required in the ZBA together with low
values \cite{Shifman} obtained in other low energy measurements have 
stimulated interest in
the light gluino hypothesis, but direct searches for such light
supersymmetric particles have, up to now, all turned out negative.  This
increases the importance of a thorough investigation of quarkonium
wave function effects beyond the ZBA.
Quarkonium has a useful analog in 
positronium \cite{positronium} although the former case
is complicated due to the non-Abelian nature of the binding force and due
to non-perturbative confinement effects.  There is continuing discussion of
the much smaller binding effects in positronium \cite{Pestieau} to which 
we can compare our
results.
     For definiteness we restrict our attention to the vector quarkonium
states such as $\Phi$, $J/\Psi$, $\Upsilon$, and in particular to the
$J/\Psi$.  In a relativistic treatment the production amplitude of
such a state with polarization $\epsilon$
together with other particles from an arbitrary initial state
takes the form
\be
     {\cal A} = \epsilon^{\ast \mu} {\cal M}_\mu .
\label{matrixel}
\ee
In the rest frame of the $J/\Psi$, the states of spin +1,-1, or zero
along the z axis (quantization axis) are described by polarization vectors
\be
\nonumber
  \epsilon^\mu(\pm)^R &=& - \frac{1}{\sqrt{2}} \left( 0 , \pm 1 , i , 0 \right)\\
  \epsilon^\mu(0)^R &=& \left( 0,0,0,1 \right) .
\label{polarizations}
\ee
(Throughout this article we use the superscript $R$ to indicate $J/\Psi$ 
rest frame
values;
for phase conventions see \cite{Gross}, page 54.)

     The production of a $J/\Psi$
is, of course, equivalent to the production
of a quark of momentum $p$ and spin $\lambda/2$ together with an antiquark
of momentum $\overline{p}$ and spin ${\overline \lambda}/2$ in a bound state
wave function $\Psi$.  The $J/\Psi$ four-momentum, $P$, and the relative
momentum, K, are 
\be
\nonumber
             P &=& p + {\overline p} \\
             K &=& (p - {\overline p})/2
\ee
with
\be
\nonumber
         P^2 &=& M^2 \\
         p^2 &=& {\overline p}^2 = m^2
\ee
Thus
\be
         K^2 &=& m^2- \frac{M^2}{4}
\label{Ksq}
\ee
The momentum, $K$, is conjugate to the relative position of the quark and
antiquark in the bound state wave function; we understand (\ref{Ksq}), 
therefore, in the sense of an average over the wave function.
     The matrix element ${\cal A}$ in (\ref{matrixel}) is therefore related
to a wave function factor, $F_w$, which is linear in $\epsilon^\ast$ and is
a matrix in the Dirac space of the heavy quarks:  
\be
    {\cal A} = \int \frac{{d^4}k}{(2 \pi)^3} \delta(\frac{P \cdot K}{M})
         {\tilde \Psi(K)}(F_w)_{\alpha \beta} {\cal O}_{\beta \alpha}.
\label{amplitude}
\ee
    Here $\alpha$ and $\beta$ are spinor indices and
$F_w$ depends on the quark and antiquark momenta and spins while
${\cal O}$ carries the dependence on other variables in the problem.  If
the initial and final states contain heavy quarks only in the bound state, then
the matrix element takes the form of a Dirac trace over $F_w$ and other
factors from ${\cal O}$.  For example, the decay amplitude for $J/\Psi$ into
an electron positron pair is
\be
{\cal A}=\int \frac{{d^4}k}{(2 \pi)^3} \delta(\frac{P \cdot K}{M}){\tilde \Psi(K)}
        \left(\mathop{\mbox{Tr}} {{\overline F}_w} \gamma_\mu \right)
     {\overline u(p^-)}\gamma^\mu v(p^+)\cdot 2 m_e \cdot 4 \pi \alpha e_q/ q^2   
\label{leptonpair}  
\ee
where $q^2$ is the virtual photon's squared four-momentum and
\be
     {\overline F}_w = \gamma_0 {F_w}^\dagger \gamma_0.
\label{Fwbar}
\ee
We use dimensionless spinors normalized to
\be
     {\overline u}u = - {\overline v}v = 1  .
\ee

   The $J/\Psi$ wave function is strongly affected by
non-perturbative confinement effects and is therefore model dependent.  In this
article we ask what can be said assuming only that the wave function is 
spherically symmetric (an S wave), that the binding is small compared to the $J/\Psi$
mass, and that the bound state can be treated as a superposition
of on-shell quark and antiquark.

    Since the production of a $J/\Psi$ is the production
of a quark-antiquark in the appropriate wave function, we would expect that, 
if $\overline \lambda = \lambda$,
for suitable quark and antiquark spinors $v$ and $\overline u$,
\be
  F_w(\lambda)_{\alpha \beta}=
\sqrt{\frac{2}{M}} \frac{2m}{\sqrt 3} 
              v({\overline p}, {\overline \lambda})_\alpha 
          {\overline u}(p,\lambda)_\beta
\label{Fwa} 
\ee
while in the case of $\overline \lambda = - \lambda$ 
we would have the spin projection zero quarkonium state
\be
     F_w(0)_{\alpha \beta}=
\sqrt{\frac{2}{M}} \frac{2m}{\sqrt 3} 
      \frac{v({\overline p}, +)_\alpha 
          {\overline u}(p,-)_\beta  + v({\overline p},-)_\alpha 
          {\overline u}(p,+)_\beta  }{\sqrt 2}.
\label{Fwb}
\ee
    It is understood that $F_w$ contains
a unit matrix in the quark color space which, with
the $\sqrt 3$ in eqs.\space\ref{Fwa} and \ref{Fwb}, puts the quark-antiquark
pair into a color singlet state. 
  The traces in eqs.\space\ref{amplitude} 
and \ref{leptonpair} include a color space trace.
In the ZBA, the normalization factors are derived in \cite{Kuehnetal} and 
\cite{BargerPhillips}.  They carry over without change to the
more general case.
Here we have used the Fourier transform and normalization identities
\be
\nonumber
\Psi(x) &=& \int \frac{d^3 K}{(2 \pi)^3} {\tilde \Psi}(K) \exp{iK \cdot x}\\
  \int d^3 x |\Psi(x)|^2 &=& 1 .
\label{Psisq}
\ee

Assuming spherical symmetry and on-shell quarks, in addition to 
(\ref{Ksq}) we have  
\be
     < {K_\mu K_\nu}> = \frac{<K ^2>}{3}\left(
   g_{\mu \nu}-\frac{P_\mu P_\nu}{M^2} \right )
\label{KmuKnu}
\ee
Binding effects in any production or decay process can, therefore, be
expressed as a power series in $<K ^2>$.  

\section{\bf The spin paradox}
\setcounter{equation}{0}

     We are free to choose any quantization axis along which to measure 
the spin of
the $J/\Psi$.  If we make the conventional choice of the z axis as quantization
axis, one might attempt to use the standard Bjorken-Drell \cite{BD} spinors
 \be
\nonumber
        u(p,\lambda)&=& \frac{1}{\sqrt{2m(p^0+m)}}(m + \mathord{\not \! p})
        \left( \begin{array}{c} \chi_\lambda \\ 0 \end{array} \right)\\
        v({\overline p},{\overline \lambda}) &=& \frac{1}
        {\sqrt{2m({\overline p}^0+m)}}
             (m - \mathord{\not \!{\overline p}})
      \left( \begin{array}{c} 0 \\ i \sigma_2 \chi_{\overline \lambda} \end{array}
           \right)
\label{uv}
\ee
with 
\be
     \chi_\lambda = \frac{(1+\lambda \sigma_3)}{2} \left( \begin{array}{c} 1 \\ 1
     \end{array} \right) .
\ee
In the case ${\overline \lambda}= \lambda$ we have
\be
\nonumber
   \left( \begin{array}{c} 0 \\ i \sigma_2 \chi_\lambda \end{array}
           \right)
    \left( \chi_\lambda \quad 0  \right)
&=& i \rho_2 (1 + \rho_3)(\lambda \sigma_1 - i \sigma_2)/4\\
   &=& {\mathord{\not \! \epsilon}}^{\ast R}(\lambda) \frac{M + {\mathord{\not \! P}}^R}{2 {\sqrt 2}M}
\label{epsR}
\ee
where the $R$ refers to the rest frame values of (\ref{polarizations}) so that 
${\mathord{\not \! P}}^R = M \gamma_0$.   In (\ref{epsR}) we have used the common
decomposition of the four-by-four Dirac matrices into two two-by-two spaces
\be
        \gamma_0 &=& \rho_3 ,\\
        \gamma^i &=& i \rho_2 \sigma^i .
\ee
Inserting (\ref{uv}) into (\ref{Fwa},\ref{Fwb}) one would obtain
\be
    (F_w)_{\alpha \beta} = \frac{1}{2M {\sqrt{3 M (p_0+m)({\overline p}_0+m)}}}
                       \left( (m - {\mathord{\not \! \overline p}}) 
                   {\mathord{\not \! \epsilon}}^{\ast R}(\lambda) (M + {\mathord{\not \! P}}^R) 
                             (m + {\mathord{\not \! p}}) I_0\right )_{\alpha \beta}   
\label{Fw2}
\ee    
where $I_0$ is the unit matrix in the three-by-three quark color space.
In general this naive use of (\ref{uv})
leads to a non-covariant $F_w$, as can be seen by
noting the explicit appearance of the $J/\Psi$ rest frame values of $\epsilon$ 
and $P$.
Problems with Lorentz covariance have been discussed by \cite{Pestieau} and 
attributed
to non-perturbative effects.
The non-covariance of (\ref{Fw2}) is connected with
a non-zero relative momentum $K$.
The rest frame values of $\mathord{\not \! \epsilon}$ are related to the values in 
a general frame
where the $J/\Psi$ has four-momentum $P^\mu = (P^0, \vec{P})$ by 
\be
       {\mathord{\not \! \epsilon}}^R = {\mathord{\not \! \epsilon}} 
              - \epsilon_0 \frac{{\mathord{\not \! P}} + {\mathord{\not \! P}}^R} {P_0 + M} .
\ee
Only if $K=0$ does (\ref{Fw2}) take the covariant form
\be
     F_w = \frac{1}{2{\sqrt {3M}}} {\mathord{\not \! \epsilon}}^{\ast}(\mathord{\not \! P} + M) .
\ee

     Although binding effects in QCD may be inherently non-perturbative,
we see this problem not as one to be resolved by assuming non-perturbative
effects, but as requiring consideration of what spinors can be covariant 
in a bound state problem.
Going beyond the ZBA requires a more careful choice of
the Dirac spinors $u$ and $v$.  Another symptom of the problem is associated 
with the way the spin of the
quarks is carried over to the spin of the bound state, as we discuss shortly.
  
    The spin operator along a
quantization axis $\tilde n$ appropriate 
to act on a Dirac spinor of momentum $p$ is
\be
      S_{\tilde n} = \frac{\mathord{\not \! p}}{2m}\gamma_5 \mathord{\not \!\tilde n} 
\ee
provided $p \cdot {\tilde n} = 0$ and ${\tilde n}^2 = -1$.  Conventionally,  
one takes  
\be
\tilde n^\mu =  (0,0,0,1) ,
\label{ntilde}
\ee
i.e., pure spacelike in the z direction.
However, this spin operator acting on the spinors of (\ref{uv}) gives 
the appropriate
$\lambda/2$ or $\overline \lambda/2$ only if
\be
      p_3 = \overline p_3 = 0 .
\ee
     In the rest frame of the bound state this implies that
\be
      K_3 =0
\ee
but, if the $J/\Psi$ has unit spin along the z axis and there is no
orbital angular momentum, then the constituent quark and antiquark must each have
a unique spin $1/2$ along this axis.  This leads to the apparent paradox that the 
relative momentum, $K$, in the $J/\Psi$ rest frame must lie in the plane
perpendicular to the quantization axis, contrary to the requirement that an s-wave
state must have a spherically symmetric distribution in the relative momentum. 

\section{\bf Recovering covariance; resolving the spin paradox}
\setcounter{equation}{0}
A free fermion spinor can only depend on the particle momentum and
the spin
axis as in (\ref{uv}).
However, in order to recover covariance in the bound state 
and resolve the spin paradox one
needs to employ not the usual Bjorken-Drell spinors (\ref{uv}) but
instead
\be
\nonumber
     u(p,\lambda) = \frac{m +\mathord{\not \! p}}{{\sqrt{2m(M/2+m)}}}\frac{M + \mathord{\not \! P}}
     {{\sqrt{2M(P_0+M)}}} 
        \left( \begin{array}{c} \chi_\lambda \\ 0 \end{array} \right)\\
     v({\overline p},{\overline \lambda}) =
     \frac{m -\mathord{\not \!{\overline p}}}{{\sqrt{2m(M/2+m)}}} \frac{M - \mathord{\not \! P}}
     {{\sqrt{2M(P_0+M)}}}
     \left( \begin{array}{c} 0 \\ i \sigma_2 \chi_{\overline \lambda} \end{array}
           \right) .
\label{uvnew}
\ee
       With these spinors in place of (\ref{uv}) the wave function factor becomes
\be
   F_w(\frac{\lambda+ {\overline \lambda}}{2}) = m \sqrt{\frac{8}{3M}}
 v({\overline p},{\overline \lambda}){\overline u}(p,\lambda) =
       \frac{(m -\mathord{\not \!{\overline p}}) \mathord{\not \! \epsilon} ^\ast (M + \mathord{\not \! P})
       (m +\mathord{\not \! p})I_0}{{\sqrt {3M}} M (M+2m)} .
\label{Fwnew}
\ee
For the spin projection zero state, the central expression is the symmetrized
combination as in (\ref{Fwb}).

     Our result agrees with that found in a recent work \cite{BodwinPetrelli}
that has appeared since we began this research.  While confirming their result,
the current article emphasizes
the covariance requirement of bound state fermion spinors depending not only 
on the fermion momentum but also on the bound state momentum.   
In addition we
consider in detail how the spin structure of the constituents carries over to the
spin of the bound state.  A comparison with other earlier work is given below.

    In calculating a quarkonium production or decay process including first order
corrections beyond the ZBA, the usual prescription is
to expand the matrix element in the relative momentum and use the expectation
values (\ref{Ksq},\ref{KmuKnu}) as we will do in the next section for
the quarkonium lepton pair decay rate and the $B$ meson to $J/\Psi + X$
rate.
The remainder of this section is devoted to
 justifying  the above choice of spinors in the quarkonium bound state.

      One should note first that the quark and antiquark spin projections
along any axis only need to sum to the quarkonium spin along that axis
after performing the integration over relative 
momentum, $K$.  For any fixed value of $K$ there can be an orbital
angular momentum that comes into the equation. 
Further, the usual spin states are defined
by reference to axes in the particle rest frame, and axis directions are not
Lorentz invariant.  Bjorken and Drell \cite{BD} specified their spinors
for nonzero fermion three-momentum in terms of the Lorentz transformation
taking $(m,0,0,0)$ to $(E, \vec{p})$, but they did not point out that such a
transformation is not unique: any rotation in the rest frame can be applied
first.  Such a rotation is the difference between the spinors we propose in
(\ref{uvnew}) and those of (\ref{uv}); such a rotation also is the
difference between a single Lorentz boost and the same boost realized via
two noncollinear boosts.
To construct
the quark and antiquark spinors with the proper spin structure we
first boost each from their differing rest frames into the rest frame of
the $J/\Psi$, since the conventional use of eqs.(\ref{polarizations}) implies
the choice of the z axis in the rest frame of
the quarkonium for the defining quantization axis.

      Any spinor in the rest frame of a particle of mass $m$ is boosted into the
 frame where this particle has four-momentum $p^\mu$ by the Lorentz
 transformation matrix
\be
         T_p = \frac{m + \mathord{\not \! p} \gamma_0}{{\sqrt{2m(m+p_0)}}} .
\label{Tp}
\ee
  Any covariant matrix, $A$, in Dirac space is boosted between these two frames by
sandwiching the matrix in the particle rest frame between $T_p$ on the left and 
\be
  {\overline T_p} = {T_p}^{-1} = \frac{m + \gamma_0 \mathord{\not \! p}}{{\sqrt{2m(m+p_0)}}}
\ee 
on the right.
The spinors (\ref{uvnew}) may be written by transforming the quark and antiquark
spinors from their rest frame, where they have spin projections $\lambda/2$ and
$\overline \lambda/2$ respectively, first to the quarkonium rest frame where they
have momenta $p^R$ and $\overline p^R$,  and then transforming to the general
frame where the quarkonium has momentum $P^\mu$.  Thus
\be
\nonumber
 u(p,\lambda) = T_P T_{p^R} \frac{1 + \lambda \gamma_5 \mathord{\not \!{\tilde n}}}{2} u_0\\
      v(\overline p,\overline \lambda) = T_P T_{\overline p^R} 
           \frac{1 + {\overline \lambda} \gamma_5 \mathord{\not \!{\tilde n}}}{2} v_0 .
\label{uvboosted}
\ee
Here,
\be
       u_0 = \left( \begin{array}{c} 1 \\ 1 \\ 0 \\ 0 \end{array} \right)
\label{u0}
\ee
\be
       v_0 = \left( \begin{array}{c} 0 \\ 0 \\ 1 \\ -1 \end{array} \right) .
\label{v0}
\ee
Note that the spinors of (\ref{uv}) are
\be
    \left( \begin{array}{c} \chi_\lambda \\ 0 \end{array} \right)&=&
      \frac{1 + \lambda \gamma_5 \mathord{\not \! {\tilde n}}}{2}u_0\\
     \left( \begin{array}{c} 0 \\ i \sigma_2 \chi_{\overline \lambda} \end{array}
           \right)&=&
      \frac{1 + \overline{\lambda} \gamma_5 \mathord{\not \! {\tilde n}}}{2}v_0  .
\ee
The quark and antiquark momenta, $p^R$ and $\overline p^R$, in the
quarkonium rest frame  are related to their values $p$ and $\overline p$ in
the general frame where the quarkonium has four-momentum $P^\mu = (P^0, \vec{P})$
by 
\be
\nonumber
      p_0^R &=& \frac{P \cdot p}{M} = \frac{M}{2}\\
      \vec{p}^R &=& \vec{p} - \frac{ \vec{P} (p_0 + M/2)}{P_0 + M}
\ee
and equivalent equations for $\overline p^R$.

  The $\gamma_0$'s in the $T_p$ and $T_{\overline p}$ of eqs. (\ref{uvboosted})
commute with $\gamma_5 \mathord{\not \! {\tilde n}}$ and
give $+1$ and $-1$ respectively acting on $u_0$ and $v_0$. We may then use
\be
\nonumber
        T_P (m + \mathord{\not \! p^R}) \overline T_P  &=& (m + \mathord{\not \! p})\\
       T_P (m - \mathord{\not \!\overline p^R}) \overline T_P  &=& (m- \mathord{\not \!\overline p})
\ee
or
\be
\nonumber
       T_P (m + \mathord{\not \!p^R}) &=& (m + \mathord{\not \! p}) T_P\\
       T_P (m - \mathord{\not \! \overline p^R}) &=& (m - \mathord{\not \!\overline p}) T_P .
\label{TPboost}
\ee
      Again, the $\gamma_0$ in $T_P$ acting on $u_0$ or $v_0$ gives $\pm 1$
respectively  so the spinors of (\ref{uvboosted}) become identical to those of
(\ref{uvnew}).    
Combining $v$ and $\overline u$ and using (\ref{epsR}) we have
\be
        F_w = m \sqrt{\frac{8}{3M}} v \overline u = 
          \frac{(m -\mathord{\not \!\overline p})}{(M/2+m)} T_P
         \frac{\mathord{\not \! \epsilon }^{\ast R} (M\! +\mathord{\not \! P^R})}{M \sqrt {12 M}} 
       \overline T_P 
        (m + \mathord{\not \! p})I_0 .
\ee
Then, using 
\be
          T_P \mathord{\not \! \epsilon }^{\ast R} (M + \mathord{\not \! P^R}) \overline T_P =\mathord{\not \!
         \epsilon }^\ast (M + \mathord{\not \! P}) ,
\ee
we have (\ref{Fwnew}).
This result agrees with that of \cite{BodwinPetrelli}.  
One might ask, however, why the conventional spinors (\ref{uv}) lead to a non-covariant result while the
spinors of (\ref{uvnew}) or (\ref{uvboosted}) lead to a satisfactory covariant result.  The answer lies in
the following observation.  A quark spinor is given by a covariant factor times a rest frame spinor.  The 
rest frame spinors combine to a covariant form in the meson rest frame (\ref{epsR}).  From there one would
have to boost to the general frame via $T_P$.  But such a spinor would not satisfy the on-shell condition.
The combination (\ref{epsR}) sandwiched with $T_{p^R}$ and ${\overline T}_{p^R}$ also provides a covariant
form in the meson rest frame.  This boosted by the $T_P$ gives a covariant form in the general frame and
also satisfies the Dirac on-shell conditions due to (\ref{TPboost}).

       To investigate how the quark and antiquark spins add to the quarkonium spin
we consider the momentum dependent spin operators for quark and antiquark
\be
\nonumber
    S_i(p) =  T_P T_{p^R} \frac{\gamma_0 \gamma_5 \gamma_i}{2} 
      \overline T_{p^R} \overline T_P
\ee

The  $S_i(p)$ satisfy the $SU(2)$ algebra
for any $p$.  
In the quark rest frame we have
\be
  \frac{\gamma_0 \gamma_5 \gamma_i}{2}= \frac{\mathord{\not \! p}}{2m}
        \gamma_5 \mathord{\not \! e_i} = \frac{\sigma_i}{2}
\ee
where $e_i$ is a pure space-like unit vector along the $i'th$ axis.
Without forcing the relative momentum to lie in the $x,y$ plane, $S_3(p)$
and $S_3(\overline p)$ have the
proper eigenvalues $\lambda/2$ and $\overline \lambda /2$ acting on the spinors of
(\ref{uvnew}).   
In addition we note that, in the quarkonium rest frame where $T_P=1$,
\be
    S_3(p) = S_3(p^R) = \frac{\mathord{\not \! p^R}}{2m}\gamma_5 \mathord{\not \! n^R} 
\ee
where
\be
       \mathord{\not \! n^R} = T_{p^R} \mathord{\not \! \tilde n} \overline T_{p^R}=
          \mathord{\not \! \tilde n} + \frac{{\tilde{\vec{n}}} \cdot \vec{p}^R}{m}
                   \frac {\mathord{\not \! p^R} + m \gamma_0}{p_0^R+m}  .
\ee
$p_0^R$, of course, is merely $M/2$.  
After averaging over $K$ directions
$\mathord{\not \! n^R}$ is in the $\mathord{\not \! \tilde n}$ direction.  
For on-shell quarks $\mathord{\not \! n^R}$ is,
after averaging, precisely $\mathord{\not \! \tilde n}$.
Thus this construction insures that
in the quarkonium rest frame the quark and antiquark spins are taken with respect
to the same $\tilde n$ direction after averaging over relative momenta.  
It is this
property that ensures covariance of the $v \overline u$ bound state and 
ensures that
the spins of the quark and antiquark carry over properly to 
the spin of the quarkonium.

  In the general frame, where the quarkonium has four-momentum $(P^0, \vec{P})$,
\be
    S_3(p) =  \frac{\mathord{\not \! p}}{2m}\gamma_5 \mathord{\not \! n} 
\ee
where
\be
       n^\mu = {n^\mu}_P - \frac{p \cdot n_P}{m} Q^\mu 
\label{nmu}
\ee
and $Q^\mu$ is the dimensionless four-vector
\be
     Q^\mu = \frac{m P^\mu + M p^\mu}{M(M/2+m)} .
\label{Qmu}
\ee
$n_P$ is the four-vector that is $\tilde n$ in the quarkonium rest frame.  
\be
      n_P = \Lambda(P) \tilde n = \left( \frac{\vec{\tilde n} \cdot 
       \vec{P}}{M},  \vec{\tilde n} + \frac{\vec{P}}{M}
                   \frac{\vec{P} \cdot \vec{\tilde n}}{(P^0+M)} \right) 
\ee
where $\Lambda(P)$ is the Lorentz transformation matrix that relates
the quarkonium rest frame to the general frame.
The four-dimensional spin axis ${\overline n}$ for the antiquark is given by
eqs.\space\ref{nmu} and \ref{Qmu} replacing $p$ by ${\overline p}$.  Only in the
$K =0$ limit or after averaging over $K$ are these axes equal.
Note that $p \cdot n = {\overline p} \cdot {\overline n} = P \cdot n_P = 0$ 
as required by Lorentz invariance.
Using these orthogonalities and the fact that
\be
     (m + \mathord{\not \! p}) \gamma_5 \mathord{\not \! Q} (M + \mathord{\not \! P}) = 0
\ee
it is easy to show that 
\be
\nonumber
      S_3(p) u(p,\lambda) &=& \frac{\lambda}{2} u(p,\lambda)\\
      S_3(\overline p) v({\overline p},{\overline \lambda}) &=& \frac{\overline 
            \lambda}{2} v({\overline p},{\overline \lambda})
\ee
where the $u$ and $v$ spinors here are those of (\ref{uvnew}).
Therefore,
\be
   S_3(\overline p)F_w + F_w S_3(p) = \frac{\lambda + \overline \lambda}{2} F_w .
\ee
This makes manifest the addition of the quark and antiquark spin projections to
compose that of the meson for arbitrary values of $K$.

    We turn now to a comparison of our result with that of other authors.
We convert all results to those appropriate to production of a quarkonium
state as opposed to a decay process and we ignore normalization factors.  
  
   In one of the seminal early papers on quarkonium processes \cite{Kuehnetal}, 
there appears the expression
\be
    F_w \sim (m - \mathord{\not \! {\overline p)}} \mathord{\not \! \epsilon ^\ast} (\gamma^0 + 1)
            (m + \mathord{\not \! p}) .
\ee
These authors did not attempt to treat the system beyond the ZBA and 
noted that the above result held in the bound state rest frame, but did not
deal with the apparent non-covariance of the result.

    In \cite{Guberinaetal}, the wave function factor was given as
\be
    F_w \sim (m - \mathord{\not \! {\overline p}}) \mathord{\not \! \epsilon ^\ast} (m + \mathord{\not \! p}) .
\label{Guberinaform}
\ee
This paper omitted the factor, $(M + \mathord{\not \! P})$, which is important in going
beyond the ZBA.  The same form appears in \cite{Petrellietal}.

     In \cite{Gross}, the form
\be
     F_w \sim \mathord{\not \! \epsilon ^\ast} (m + \mathord{\not \! P})
\ee
is given without attempting an analysis beyond the ZBA.

    In \cite{Pestieau}, in addition to the form (\ref{Guberinaform}), there 
appears the non-covariant form
\be
    F_w \sim \mathord{\not \! \epsilon ^\ast }(\gamma^0 + 1).
\ee
Again, these authors did not deal with the apparent non-covariance, preferring 
to restrict their considerations to the meson rest frame.

    Repeating our analysis above, it is straightforward to write the wave function
factor for a bound state of different quark species.  For a quark of mass $m$ and
antiquark of mass $\overline m$, the appropriate generalization is
\be
  F_w(\lambda) = \frac{({\overline m}-\mathord{\not \!{\overline p}})\mathord{\not \! 
          \epsilon ^\ast} 
        (M + \mathord{\not \! P})(m +\mathord{\not \! p})I_0}
        {M \sqrt{ 12 M({p_0}^R+m)({{\overline p}_0}^R + \overline{m})}} .
\ee
Here the quark and antiquark energies in the rest frame of the bound state are
\be
\nonumber
      {p_0}^R = \frac{M}{2} + \frac{m^2 - {\overline m}^2}{2M}\\
   {{\overline p}_0}^R = \frac{M}{2} -\frac{m^2 - {\overline m}^2}{2M}  .
\ee
In the case of non-equal masses, the $\delta$ function in
(\ref{amplitude}) is modified to
\be
     \delta(\frac{P \cdot K}{M} - \frac{m^2 - {\overline m}^2}{M})
\ee

\section{\bf Decay rates beyond the zero-binding approximation}
\setcounter{equation}{0}

     We would now like to use the bound state formalism to calculate the
decay rates $J/\Psi \rightarrow l^+ l^-$ and $B \rightarrow J/\Psi + X$ including
the first corrections beyond the ZBA.
     In the decay of a $J/\Psi$ as opposed to the production one uses the 
conjugate wave function factor (\ref{Fwbar}).
Using (\ref{Fwa}) in the trace of (\ref{leptonpair}) and 
averaging over the relative
momenta using (\ref{Ksq},\ref{KmuKnu}) we have
\be
   \mbox{Tr} ({\overline F}_w \gamma_\mu) = 
     \frac{{\sqrt 3}(2 m + M)}{\sqrt M} \epsilon_\mu \left(1 - \frac{4K^2}
       {3(2m+M)^2}
       \right) \sim \sqrt{12}\epsilon_\mu \left( 1 +\frac{2K^2}{3 M^2}\right) .
\ee
Thus, to lowest order in $K ^2$,
\be
    \Gamma(J/\Psi \rightarrow l^+ l^-) = 16 \pi \alpha^2 {e_q}^2 
             \frac{{| \Psi(0) |}^2}{M^2}\left( 1 + \frac{4K^2}{3M^2} \right) .
\ee
Using the Particle Data Group \cite{PDG} charm quark mass 
$m=(1.25\pm 0.1)$ GeV and $J/\Psi$ mass
$M=3.097$ GeV we see that the first order correction beyond the
ZBA decreases the leptonic pair decay rate of the $J/\Psi$ by a
factor of $0.88 \pm .04$.  
This correction has been also found in \cite{Pestieau}.
The effect goes in the wrong direction to explain the 
suppression
of the hadronic $J/\Psi$ decays relative to the leptonic pair decay. 
However, important
contributions come from the binding corrections to the
three gluon decay.  Although the first order correction does
reduce the three gluon decay rate, the authors of 
\cite{BodwinPetrelli} note an apparent slow
convergence of the $K^2$ expansion indicating that further study is needed.

     Turning to the decay $B \rightarrow J/\Psi + X$, the Dirac structure of the
invariant amplitude is
\be
    {\cal A} \sim {\overline u}(p_s) \gamma^\mu P_L F_w \gamma^\nu P_L u(p_b)
     \left(-g_{\mu \nu}+ \frac{k_\mu k_\nu}{{M_W}^2}\right) \frac{1}{{M_W}^2-k^2}
\ee
where $P_L$ is the left handed chiral projection and
\be
     k_\mu = p_{b \mu}-p_\mu = p_{s \mu}+\overline{p}_\mu .
\ee
It is easy to see that,
to lowest order in $\frac{1}{{M_W}^2}$ and to first order in
$\frac{K^2}{M^2}$, the decay rate is
modified by the identical factor $(1 + \frac{4K^2}{3M^2})$ that occurs in 
lepton pair decay. Thus, measured relative to lepton pair decay
or, equivalently, using the leptonic rate as a measure of $|\Psi(0)|^2$, the 
relativistic
corrections do not change the inclusive $J/\Psi$ decay rate of the $B$ meson at
zeroth order in the strong coupling constant.

    In summary, we have shown how the recovery of Lorentz covariance requires the
use of quark spinors in the bound state that depend not only on the quark 
momentum but also on the bound state momentum.  We have also noted how the 
relevant spin operators for quark and antiquark refer to momentum dependent 
axes that are only equal in an average sense. We have confirmed earlier recent 
results for the wave
function factor and lepton-pair decay mode of quarkonium, and have extended the
beyond-ZBA analyses to the inclusive $J/\Psi$ decay of the $B$ meson.

{\bf Acknowledgements}

    This work was supported in part by the US Department of Energy under grant
DE-FG02-96ER-40967.  We thank Philip Coulter for discussions in the early 
stages of
this research.

\end{document}